\documentclass{jnmp}

\usepackage{amsmath}

\JNMPnumberwithin{equation}{section}

\begin{document}

\renewcommand{\evenhead}{H L Morrison and A D Speliotopoulos}
\renewcommand{\oddhead}{The Landau--Ginzberg Theory for the Two-Dimensional Bose Gas}

\thispagestyle{empty}

\FirstPageHead{9}{4}{2002}{\pageref{Morrison-firstpage}--\pageref{Morrison-lastpage}}{Article}

\copyrightnote{2002}{H L Morrison and A D Speliotopoulos}

\Name{The Landau--Ginzberg Theory\\ for the Two-Dimensional Bose Gas}
\label{Morrison-firstpage}

\Author{Harry L MORRISON~$^\dag$ and Achilles D SPELIOTOPOULOS~$^\ddag$}

\Address{$^\dag$ Department of Physics, University of California, Berkeley,
California 94720, USA\\[10pt]
$^\ddag$ National Research Council, Board on Physics and Astronomy,\\
~~2101 Constitution Avenue, NW, Washington, DC 20418\\
~~Current address: Department of Physics, University of California, Berkeley, CA, 94720-7300\\
~~E-mail: adpelio@uclink.berkeley.edu}

\Date{Received December 24, 2001;
Accepted May 29, 2002}

\begin{abstract}
\noindent
Using results from sheaf theory combined with the phenomenological
theory of the two-dimensional superfluid, the precipitation of quantum
vortices is shown to be the genesis of a macroscopic order parameter for a
phase transition in two dimensions.
\end{abstract}

\section{Introduction}

The underlying mechanism for superfluidity in three-dimensional Bose
systems is based upon the spectral decomposition of the Bose field
operator $\Psi(x)$. The zero-momentum component of the operator
becomes the macroscopic order parameter for the associated phase
transition~\cite{Haag} in the thermodynamic limit. The existence of
this order parameter, which defines the condensed phase~\cite{FW}, is a
consequence of the breaking of a $U(1)$ gauge symmetry. Indeed, the
states of the condensate and the states of the normal fluid belong to
two inequivalent representations of the algebra generated by
$\Psi(x)$ and $\Psi^\dagger(x)$.

In two-dimensions, however, the situation is much more complicated
since the breaking of a $U(1)$ gauge symmetry is explicitly prohibited
\cite{Hoh,MW,Mer,GWM}. There are
instead two phenomenological theories of the phase transition, both of
which presupposes the existence of vortices with integral vorticity, or
``charge'', in the system. In the Kosterlitz--Thouless--Nelson
\cite{KT,NK,Nel,Minn} theory the
superfluid transition is marked by the precipitation of essentially
free vortices in the normal fluid into vortices that are tightly bound
in dipole pairs in the superfluid. There are effectively no free
vortices in the superfluid state in this theory. In theory given in~\cite{SM},
on the other hand, vortices are only present in the
superfluid state and the phase transition is due to the annihilation
of oppositely charged vortices at the transition temperature. In this
case, there are no {\it quantum\/} vortices in the normal fluid
state. An essential feature of both theories, however, is that the net
vorticity, or charge, of the system must be zero. Indeed, it was shown
in~\cite{SM} that this is a necessary condition for the phase
transition to occur.

In this paper we use the existence of quantized vortices in the
superfluid state to demonstrate by construction the existence of an
order parameter for the superfluid phase transition in spite of the
absence of a Bose-condensation-based phase transition in
two-dimensions. We do so by using the results from analysing the
classical equations of motion for the vortices given in~\cite{KT,NK,Nel,Minn}
and~\cite{SM} as a motivation for studying sheaf theory. Results from sheaf theory are
then combined with an effective-field description of the system to
construct the order parameter. Global sheaf-theoretic results on the
geometry of two-dimensional surfaces are then used to make statements
about the possibility of having the superfluid phase transitions on
various two-dimensional surfaces.

The rest of this paper is organized as follows. In Section~2 we
review the properties of the vortex gas used by both phenomenological
theories of the superfluid phase transition. Results from this section
is meant to motive the use of sheaf theory as a tool for constructing
the order parameter for the phase transition. In Section~3 we review
those aspects of sheaf theory that are needed for this construction. The
results stated in this section are all well known and no proofs are
given. In Section~4 we combine the results of Section~3 with an
effective free energy functional to construct the order parameter for
the phase transition. Results from sheaf theory will also be used
to comment on the existence of the superfluid phase transition for
various two-dimensional surfaces. Final concluding remarks can be
found in Section~5.

\section{Vortices in two-dimensions}

The description of the two-dimensional quantum vortices used in
\cite{KT,NK,Nel,Minn} and~\cite{SM}
follows in direct analogy with the description of vortex lines used in
three-dimensional superfluids. In this description vortices are
characterized by $\{\kappa_i, z_i\}$, where $\kappa_i$ is the
circulation associated with the $i^{\rm th}$ vortex at the position
$z_i$ (in complex coordinates) on a Riemann surface $\mathbf M$. Vortices
are formed due to a current flow $j_z$ in the fluid. Because of the
underlying quantum  mechanical nature of the superfluid, their circulation
(or vorticity) is quantized and
\begin{equation}
\kappa_i \equiv \int_{\gamma_i}\frac{j_{\bar z} dz}{\rho} =
\frac{h}{m} n_i,
\label{e1}
\end{equation}
where $\rho$ is the (constant) superfluid density, $\gamma_i$ is a
closed contour on $\mathbf M$ encompassing the $i^{\rm th}$ vortex and
$n_i$ is a non-zero integer.

One then traditionally appeals to classical fluid dynamical arguments
\cite{Fried} to describe the interaction between vortices and their
dynamics. Because of the particular properties of two
dimensional fluid flow, one can treat the vortices as though they
were particles in and of themselves with a hamiltonian
\begin{equation}
H = - \frac{\hbar^2}{m^2}\sum_{k>l} n_k n_l \log|z_k-z_l|.
\label{e2}
\end{equation}
This hamiltonian is identical to the interaction hamiltonian of a gas
of charges particles with charge $n_i$, hence the analogy between
a gas of vortices and a gas of charged particles. Accordingly, $n_k$
is often interpreted as the charge of the $k^{\rm th}$ vortex, and
the total charge of the vortices is
\begin{equation}
Q = \sum_k n_k.
\label{e3}
\end{equation}
The evolution equations for vortices is quite different from that of
charged particles, however, \cite{Fried,RM,MAL,SM2} in that
\begin{gather}
n_k\frac{dz_k}{dt} = -2i\frac{m}{\hbar}\frac{\partial H}
            {\partial \bar z_k},
\nonumber \\
n_k\frac{d\bar z_k}{dt} = 2i\frac{m}{\hbar}\frac{\partial
            H}{\partial z_k}.
\label{e4}
\end{gather}
Notice that vortex motion is determined by its {\it velocity}, and not
its acceleration.

It is straightforward to see that $H$ admits the following affine
transformation in the complex plane
\begin{equation}
H(z_1, \dots, z_n) = H(\eta z_1+\xi, \dots, \eta z_n+\xi),
\label{e5}
\end{equation}
where $\xi$ is an uniform time dependent translation, and $\eta$ (with
the constraint that $|\eta|=1$) is a time dependent uniform
rotation. Translational invariance gives the constant of the motion
\begin{equation}
M = \sum_k n_k z_k,
\label{e6}
\end{equation}
while rotational invariance gives
\begin{equation}
I = \sum_k n_k | z_k|^2.
\label{e7}
\end{equation}

To motive our use of sheaf theory to study this system, we associate
to each $M$ a~meromorphic function
\begin{equation}
f(z)= \prod_k (z-z_k)^{n_k},
\label{e8}
\end{equation}
with poles or zeros of order $n_k$ at the points $z_k$ on $\mathbf
M$. Such a function is often identified with the flow potential of an
ideal fluid. This  association is clearly not unique, of course, for
if $g(z)$ is any non-vanishing holomorphic function on an open set
$U\subset {\mathbf M}$, then $f(z)g(z)$ can also be associated to the same
$M$; $g$ does not add any zeros or poles to $f$. We should instead
consider equivalence classes of meromorphic functions in which two
meromorphic functions $f(z)$ and $h(z)$, not identically zero, are
equivalent if $f/h$ is a non-vanishing holomorphic function. This
leads naturally to the consideration of all such equivalence classes
over the manifold ${\mathbf M}$. In sheaf theory this equivalence class is
the sheaf of germs of divisors with $M$ identified with $\mathfrak{d}$, the
divisor of a line bundle over $\mathbf M$. Correspondingly, $Q$~is
identified with the bundle's chern class. Sheaf theory is therefore a
natural framework for describing and understanding the behavior and
dynamics of quantum vortices.

\section{Overview of sheaf theory}

In this section we present a brief review of sheaf theory, highlighting
the topics needed in our analysis. We follow exclusively the treatment
and notation found in~\cite{Gun} (see also~\cite{Grif}). No new
results in sheaf theory is presented, however, and we do not present
any proofs for the results stated here.

We begin with with an open covering $\mathcal U$ of a Riemann surface $\mathbf M$
consisting of open sets $\{ U_\alpha\}$. As an additional structure,
the notion of a sheaf is introduced:

A {\it sheaf} of abelian groups over a topological space $\mathbf
M$ is a topological space $\mathcal S$, together with a mapping
$\pi : {\mathcal S}\to{\mathbf M}$ such that:

\begin{enumerate}
\vspace{-2mm}
\itemsep0mm
\item[1.] $\pi$ is a local homomorphism;

\item[2.] for each point $p\in {\mathbf M}$, the set ${\mathcal S}_p =
             \pi^{-1}(p)$, ${\mathcal S}_p \supset {\mathcal S}$ has the structure
         of an abelian group;

\item[3.] the group operations are continuous in the topology of $\mathcal S$.
\vspace{-2mm}
\end{enumerate}

${\mathcal S}_p$ is called the {\it stalk} of the sheaf $\mathcal S$ at the
point $p$. A {\it section} $s$ of the sheaf $\mathcal S$ over $U_\alpha$
is a continuous map $s: U_\alpha\to {\mathcal S}$ such that $\pi \circ s
: U_\alpha \to U_\alpha$ is the identity map on $\mathbf
M$. $\Gamma(U_\alpha, {\mathcal S})$ is the set of all sections of $\mathcal
S$ over $U_\alpha$ and $\Gamma({\mathcal U}, {\mathcal S})$ is the formal sum
of all $\Gamma(U_\alpha,{\mathcal S})$ over $\mathcal U$.

If $f$ and $g$ are any two functions defined on the neighborhoods
$U_\alpha$ and $U_\beta$ respectively such that $U_\alpha \cap
U_\beta\not= \emptyset$, then $f$ and $g$ are equivalent at a
point $p\in U_\alpha \cap U_\beta$ if $f(z)=g(z)$ for all $z$ in a
neighborhood of $p$ contained in $U_\alpha \cap U_\beta$. The {\it
germ} of a function at $p$ is the equivalence class of local
functions at $p$ and is denoted by $f_p$. The set of all germs of
functions at each point $p$ in $\mathbf M$ forms an abelian groups
under addition and multiplication. As such the following sheaves
can be defined:

\medskip

${\mathcal C}^\infty$, {\it the sheaf of germs of infinitely
differentiable functions};

$\vartheta$, {\it the sheaf of germs of holomorphic functions};

$\vartheta^{*}$, {\it the sheaf of germs of non-vanishing
holomorphic functions};

${\mathcal M}$, {\it the sheaf of germs of meromorphic functions};

${\mathcal M}^{*}$, {\it the sheaf of germs of non-vanishing meromorphic
functions}.

\medskip

These sheaves may be ordered by:
\begin{equation}
{\mathcal C}^\infty\supset{\mathcal M}\supset{\mathcal M}^{*}\supset
\vartheta\supset\vartheta^{*},
\label{e9}
\end{equation}
In addition, the quotient sheaf $\mathbb{D}\equiv {\mathcal
M}^{*}/{\vartheta^{*}}$ is the {\it sheaf of divisors} and is of
principle interest to us.

Consider next a function  $f\in {\mathcal M}^{*}$ defined on
$U_\alpha$. The order of $f$ at a point $p\in U_\alpha$, denoted by
$\nu_p(f)$, is the order of the first non-zero coefficient of the
Laurent expansion of $f$ about the point $z=p$. Furthermore, for
each $f,g\in {\mathcal M}^{*}$, $\nu_p(fg) = \nu_p(f) + \nu_p(g)$. If $g$
is also an element of $\vartheta^{*}$, then $\nu_p(g)=0$ and
$\nu_p(gf)=\nu_p(f)$. The {\it divisor} is the quotient map
$\mathfrak{d}:{\mathcal M}^{*}\to \mathbb{D}$ defined as
\begin{equation}
\mathfrak{d}(f) \equiv \sum_p \nu_p(f) p,
\label{e10}
\end{equation}
and there is an exact sequence of sheaves
\begin{equation}
O \; \to \; \vartheta^{*} \; \frac{\iota}{\to}\; {\mathcal
M}^{*}\; \frac{\mathfrak{d}}{\to}\; \mathbb{D}\; \to \; O,
\label{e11}
\end{equation}
where $\iota$ is the inclusion map and $O$ is the trivial sheaf. A
sheaf is trivial if for each $p\in {\mathbf M}$, $O$ is the
trivial group.

Cohomology classes of a sheaf $\mathcal S$ are defined in the
usual way. To each open covering $\mathcal U$ of $\mathbf M$ is
associated a simplicial complex $N({\mathcal U})$ called the nerve
of $\mathcal U$ with each vertex of $N({\mathcal U})$ contained in
one $U_\alpha$. The vertices $U_0, \dots, U_q$ span a $q$-complex
$\sigma = (U_0,\dots, U_q)$ if and only if $U_0 \cap \dots\cap
U_q\not=\emptyset$. $| \sigma| =U_0 \cap \dots\cap U_q$ is
the support of $\sigma$. For any sheaf of abelian groups $\mathcal
S$, a $q$-cochain of $\mathcal U$ with coefficients in the sheaf
$\mathcal S$ is a function $f$ which associates every $q$-simplex
$\sigma\in N({\mathcal U})$ a section
$f(\sigma)\in\Gamma(|\sigma|,{\mathcal S})$. The set of all
$q$-chains is $C^q({\mathcal U}, {\mathcal S})$. Since $f,g\in
C^q({\mathcal U}, {\mathcal S})\Rightarrow f+g \in C^q({\mathcal
U}, {\mathcal S})$, $C^q({\mathcal U}, {\mathcal S})$ forms an
abelian group.

The coboundary operator $\delta:C^q({\mathcal U}, {\mathcal S})\to
C^{q+1}({\mathcal U}, {\mathcal S})$ is defined as follows. For $f\in
C^q({\mathcal U},{\mathcal S})$, and a $q+1$-simplex $q=(U_0,\dots. U_{q+1})\in
N({\mathcal U})$,
\begin{equation}
(\delta f)(U_0, \dots, U_{q+1}) = \sum_{i=0}^{q+1}(-1)^i \rho_{|\sigma|}
f(U_0,\dots, U_{i-1}, U_{i+1}, \dots, U_{q+1}),
\label{e12}
\end{equation}
where $\rho_{|\sigma|}$ is the restriction of $f(U_0,\dots,
U_{i-1}, U_{i+1}, \dots, U_{q+1})\in \Gamma(U_0\cap\dots\cap
U_{i-1}\cap U_{i+1}\cap\dots \cap U_{q+1})$ to
$|\sigma|$. $\delta$ is a group homomorphism where
$\delta^2=0$. The subset $Z^q({\mathcal U}, {\mathcal S})=\{f\in C^q({\mathcal
U}, {\mathcal S}) \, | \,  \delta f=0\}$ is the group of $q$-cocycles while
the image $\delta C^{q-1}({\mathcal U},{\mathcal S})$ is the group of
$q$-coboundaries. Since $C^q({\mathcal U}, {\mathcal S})\subset \delta
C^{q-1}({\mathcal U}, {\mathcal S})$ and $Z^q({\mathcal U}, {\mathcal S})\subset
\delta C^{q-1}({\mathcal U}, {\mathcal S})$, the quotient group for $q>0$ is
\begin{equation}
H^q({\mathcal U},{\mathcal S}) = Z^q({\mathcal U}, {\mathcal S})/\delta C^{q-1}({\mathcal
U}, {\mathcal S}),
\label{e13}
\end{equation}
while $H^0=Z^0({\mathcal U}, {\mathcal S})$. This is the $q^{\rm th}$
cohomology group of ${\mathcal U}$ with coefficients in the
sheaf~$\mathcal S$. Although this definition of $H^q({\mathcal U}, {\mathcal S})$ depends
explicitly on the choice of covering $\mathcal U$ of $\mathbf M$, the
following limit~\cite{Gun},
\begin{equation}
H^q({\mathbf M},{\mathcal S}) = {\rm dir.~lim.~}_{\mathcal U} H^q({\mathcal U}, {\mathcal S}),
\label{e14}
\end{equation}
gives a $H^q({\mathbf M}, {\mathcal S})$ that is independent of any specific
choice of $\mathcal U$.

It is known that for any paracompact space $\mathbf M$, $H^q({\mathbf M},
{\mathcal S})=0$. Furthermore, if $\mathbf M$ is also non-compact then
$H^q({\mathbf M},\vartheta)=0$ for all $q>0$.

For a compact manifold, on the other hand, $H^0({\mathbf M},{\mathcal
S})=\Gamma({\mathbf M},\vartheta) = C$, while $H^q({\mathbf M}, \vartheta)$ $=0$
for $q\ge 2$. The only non-trivial cohomology class remaining is
\begin{equation}
H^1({\mathbf M},\vartheta) \approx \Gamma({\mathbf M}, {\mathcal
C}^\infty)\Big/\frac{\partial\Gamma({\mathbf M}, {\mathcal C}^\infty)}{\partial z}.
\label{e15}
\end{equation}
Furthermore, $H^1({\mathbf M}, {\mathcal M})=0$, which leads to the
fundamental existence theorem of Riemann surfaces: every line bundle
on $\mathbf M$ has a non-trivial meromorphic cross-section, hence every
line bundle is the bundle of a divisor. Instead of studying the
structure of $\mathfrak{d}$, we need only consider the structure of complex
line bundles of $\mathbf M$.

To study the structure of $\mathbf M$ itself, we consider the subgroup
$H^1({\mathbf M}, \vartheta^{*})$, the group of complex line bundles over
${\mathbf M}$. For every $\xi\in H^1({\mathbf M},\vartheta^{*})$, choose a basis
${\mathcal U}=\{U_\alpha\}$ of the open covering of $\mathbf M$ and a cocycle
$(\xi_{\alpha\beta})\in Z^1({\mathbf
M},\vartheta^{*})$. $\xi_{\alpha\beta}$ is a holomorphic, nowhere
vanishing function on $U_\alpha\cap U_\beta$. The cocycle condition
insures that for $p\in U_\alpha\cap U_\beta\cap U_\gamma$,
$\xi_{\alpha\beta}(p)\cdot \xi_{\beta\gamma}(p) =
\xi_{\alpha\gamma}(p)$. For each $U_\alpha$ the group ${\mathcal S}_\alpha
= \Gamma(U_\alpha,\vartheta)$ is associated with a group homomorphism
$\tau_{\beta\alpha} : {\mathcal S}_\alpha \to {\mathcal S}_\beta$ defined on
the inclusion $U_\alpha \supset U_\beta$ by:
\begin{equation}
(\tau_{\beta\alpha} f)(p) = \xi_{\beta\alpha}\cdot f(p),
\label{e16}
\end{equation}
for $p\in U_\alpha\subset U_\beta$, $f\in {\mathcal S}_\alpha$ and
$\tau_{\beta\alpha}(f) \in {\mathcal S}_\beta$. On the triple overlap
$U_\alpha \subset U_\beta \subset U_\gamma$ and for $f\in {\mathcal S}_\alpha$,
$(\tau_{\gamma\beta}(\tau_{\beta\alpha}(f)))(p) =
(\tau_{\gamma\alpha}f)(p)$ for all $p\in U_\gamma$. $\{ {\mathcal U},
{\mathcal S}_\alpha, \tau_{\alpha\beta}\}$ forms a pre-sheaf and the
associated sheaf $\vartheta(\xi)$ is the sheaf of germs of holomorphic
cross-sections of the line bundle $\xi$. Similarly, taking ${\mathcal
S}_\alpha=\Gamma({\mathcal U}_\alpha, {\mathcal M}^{*})$ results in ${\mathcal
M}^{*}(\xi)$, the sheaf of germs of not identically vanishing
meromorphic cross-sections of the line bundle $\xi$. Clearly for
$\xi = 1$ we recover $\vartheta(1) =\vartheta$ and ${\mathcal M}^{*}(1) =
{\mathcal M}^{*}$.

Among all the possible bundles formed from members of $H^1({\mathbf M},
\vartheta^{*})$ there is the canonical bundle $\kappa$ which contains
information on the structure of $\mathbf M$ itself. Because $\mathbf
M$ is a Riemann surface, given an open covering $\mathcal U$ of $\mathbf M$
and charts $z_\alpha$ defined on $U_\alpha$, there are local
holomorphic transition functions $f_{\alpha\beta}:z_\alpha\to z_\beta$
where $z_\alpha(p)= f_{\alpha\beta}(z_\beta(p))$ for all $p\in
U_\alpha\cap U_\beta$. These transition functions form the canonical
bundle by defining
\begin{equation}
\kappa_{\alpha\beta}(p) = \left[f_{\alpha\beta}'(z_\beta(p))\right]^{-1},
\label{e17}
\end{equation}
For $p\in U_\alpha\cap U_\beta\cap U_\gamma$,
$z_\alpha(p)=f_{\alpha\beta}(f_{\beta\gamma}(z_\gamma(p))$, and using the
chain rule,
\begin{equation}
\kappa_{\alpha\gamma}(p) \equiv
\left[f_{\alpha\gamma}'(z_\gamma(p))\right]^{-1}=
\left[f_{\alpha\beta}'(f_{\beta\gamma}(z_\gamma(p)))\right]^{-1}
\left[f_{\beta\gamma}'(z_\gamma(p))\right]^{-1}=\kappa_{\alpha\beta}(p)
\kappa_{\beta\gamma}(p)
\label{e18}
\end{equation}
$\kappa$ satisfies the cocycle condition. Clearly $\{\kappa_{\alpha\beta}\}$
are just the transitions functions of the tangent space of $\mathbf M$.

The canonical bundle is used to study the geometry of $\mathbf M$
by considering the exact sequence of sheaves
\begin{equation}
{\mathcal O}\; \to \; {\mathbf Z}\; \to \; \vartheta \; \frac{e}{\to} \; \vartheta^{*}\; \to \; {\mathcal O},
\label{e19}
\end{equation}
with $e(f)\equiv exp(2\pi i f)$ for $f\in \vartheta$. Corresponding to
this sequence there is the exact cohomology sequence
\begin{equation}
{\mathcal O}\; \to \; H^1({\mathbf M},\vartheta)/H^1({\mathbf M}, {\mathbf Z})\; \to \;
H^1({\mathbf M},\vartheta^{*}) \; \to \; H^2({\mathbf M},{\mathbf Z})\; \to \; {\mathcal O},
\label{e20}
\end{equation}
$c:H^1({\mathbf M}, \vartheta^{*})\to H^2({\mathbf M},{\mathbf Z})$ is the
characteristic homomorphism and for each line bundle $\xi\in H^1({\mathbf
M}, \vartheta^{*})$ the image $c(\xi)$ is the chern class of the line
bundle $\xi$. A specific representation of $c(\xi)$ can be formed by
taking considering the $C^\infty$ function $r_\alpha$ defined on
$U_\alpha$ such that for $p\in U_\alpha\cap U_\beta$,
$r_\alpha(p)=r_\beta(p)|\xi_{\alpha\beta}|^2$. Then
\begin{equation}
c(\xi) = \frac{1}{2\pi i}\int_{\mathbf M} \frac{\partial}{\partial z}
\frac{\partial}{\partial \bar z}\log r_\alpha \; dz d\bar z,
\label{e21}
\end{equation}
Using this, one finds that for any function $f\in \Gamma({\mathbf M},
{\mathcal M}^{*})$, the chern class of that bundle will be
\begin{equation}
c(\xi) = \sum_{p\in {\mathbf M}}\nu_p(f),\qquad {\rm where}\qquad
\mathfrak{d}(f)= \sum_{p\in {\mathbf M}}\nu_p(f)p,
\label{e22}
\end{equation}
It is important to note that the chern class of the bundle will have
the same value for {\it every} function $f$ in $\Gamma({\mathbf M}, {\mathcal
M}^{*}(\xi))$.

As we shall see in the next section, the chern class of the bundle
$\xi$ is identified with the net charge of the vortices. Because we
are primarily interested in the superfluid phase transition, we shall
restrict our attention to bundles whose chern class vanishes. Referring
back to the exact cohomology sequence, we find that the subset of
$H^1({\mathbf M},\vartheta^{*})$ with vanishing
chern class is
\begin{equation}
\{
    \xi\in H^1({\mathbf M},\vartheta^{*}) | c(\xi)=0
                                             \} \approx
     \frac{H^1({\mathbf M},\vartheta)}{H^1({\mathbf M},Z)}
     \approx
     \frac {H^1({\mathbf M},{\mathbf C})}
          {
        H^1({\mathbf M},{\mathbf Z})+\delta\Gamma({\mathbf M},\vartheta^{1,0})
      }
\label{e23}
\end{equation}
where $\vartheta^{1,0}$ is the sheaf of germs of holomorphic
1-forms. The line bundles whose chern class vanishes are precisely
those whose representative cocycles $\{\xi_{\alpha\beta}\}$ consists
only of constant functions.

Finally, we note that from the Riemann--Roch theorem, dim
$H^1({\mathbf M},{\mathbf C})=2g$ where $g$ is the genus of the
surface $\mathbf M$. A $g=0$ surface is a sphere while a $g=1$
surface is the complex torus. It is known that $H^1({\mathbf M},
{\mathbf Z})$ forms a lattice subgroup of $H^1({\mathbf
M},{\mathbf C})$. This means that any $2g$ generators of
$H^1({\mathbf M}, {\mathbf Z})$ forms a basis for $H^1({\mathbf
M},{\mathbf C})$. The quotient group $H^1({\mathbf M},{\mathbf
C})/H^1({\mathbf M},{\mathbf Z})=({\mathbf R}/{\mathbf Z})^{2g}$,
which is the $g$-torus. In fact, $({\mathbf R}/{\mathbf Z})^{2g}$
forms an abelian complex Lie group. The subgroup of $H^1({\mathbf
M},\vartheta^{*})$ with vanishing chern class is isomorphic to an
abelian complex Lie group.

\section{Application of sheaf theory}

We have seen in the previous sections that sheaf theory provides a
natural framework for describing and analyzing the behavior of
vortices in two dimensional superfluids. In this section we will apply
the results of sheaf theory and structure of the phenomenological
theories of the superfluid phase transition in two dimensions to
construct an order parameter for the transition. To do so we begin by
identifying the configuration space of the Bose liquid with the
Riemann surface $\mathbf M$. Because all the cohomology groups of a
paracompact manifold vanishes, for vortices to be present $\mathbf M$ must
be a compact manifold. Then at each time $t$ there is a specific
configuration of vortices and a corresponding constant of the motion
$M$ identified with the divisor $\mathfrak{d}$ of an equivalence class of
functions $f\in {\mathcal M}^{*}$. The net charge of the system is
identified with the chern class of the line bundle $\xi$, which is
in turn interpreted as the internal symmetry space for the
system. Since $\xi$ must be constants, the symmetry group must be
global and, as we shall see later, will be of the form $U(1){\mathrm
X}\cdots {\mathrm X}U(1)$.

This mathematical structure is rich enough to encompass any
configuration of vortices. Every member of ${\mathcal M}^{*}$ has the same
chern class; ${\mathcal M}^{*}$ thus contains functions representing not
only all possible positions of vortices, but also all possible number
of vortices. As yet, $\xi$ is not specific enough to single out any
one specific configuration of vortices. It is treated instead as a
parameter that will determine the existence, and, perhaps, genesis of
vortices. As with all physical parameters $\xi$ should arise from the
partition function
\begin{equation}
{\mathcal Z} = \int_{\Gamma({\mathbf M},{\mathbf C}^\infty(\xi))}
{\mathbf\it d}\Psi{\mathbf\it d}\Psi^\dagger\exp\left(-\beta\int_{\mathbf M}
F(\Psi,\Psi^\dagger)d^2x\right)
\label{e24}
\end{equation}
where the functional integration is over are sections $\Gamma({\mathbf
M},{\mathbf C}^\infty(\xi))$ of the sheaf $C^\infty(\xi)$ and $F$ is the
free energy functional of an effective field $\Psi\in \Gamma({\mathbf M},
{\mathbf C}^\infty(\xi))$ for the system. The difficulty in determining
$\xi$ lies in the observation that the domain of the functional
integration and, to a lesser extent, $\mathbf M$ itself, is determined by
$\xi$. To find $\mathcal Z$, we must find $\xi$. Unfortunately, to find
$\xi$, we need $\mathcal Z$. To decouple the system, we expand the free
energy about its minimum
\begin{gather}
{\mathcal Z} = \exp\left(-\beta\int_{\mathbf M}F_{\min}d^2x\right)
      \int_{\Gamma({\mathbf M},{\mathbf C}^\infty(\xi))}
          {\mathcal D}\Psi{\mathcal D}\Psi^\dagger
\nonumber \\
\phantom{{\mathcal Z} =}{}\times
          \exp\Bigg((-\beta\int_{\mathbf M}
        \Bigg(\frac{1}{2}\frac{\delta^2F}{\delta\Psi^2}
            \Bigg|_{\Psi_{\min}}(\Psi-\Psi_{\min})^2+
                     \frac{\delta^2F}{\delta\Psi\delta\Psi^\dagger}
            \Bigg|_{\Psi_{\min}} | (\Psi-\Psi_{\min})|^2
             \nonumber \\
\phantom{{\mathcal Z} =}{}+
             \frac{1}{2}\frac{\delta^2F}{\delta{\Psi^\dagger}^2}
            \Bigg|_{\Psi_{\min}}
            (\Psi^\dagger-\Psi_{\min}^\dagger)^2+\cdots\Bigg)
          d^2x\Bigg). \label{e25}
\end{gather}
where $\Psi_{\min}$ is determined by the solution of
\begin{equation}
\frac{\delta F}{\delta \Psi}\Bigg|_{\Psi_{\min}} = 0,
\label{e26}
\end{equation}
and is the order parameter.

In principle, the exact form of $F$ can be determined by the
microscopic, or ``bare'' fields $\Psi_{\rm bare}$ and the corresponding
microscopic Hamiltonian. When the strong interaction between helium
atoms is taken into account, $\Psi_{\rm bare}$ is replaced by the effective
field $\Psi$ that takes into account interactions. Consequently, the
detailed form of $F$ can be very complicated. Nevertheless, we can
expand $F$ in a power series in $|\Psi|^2$,
\begin{equation}
F(\Psi,\Psi^\dagger)=\frac{1}{2m}|\nabla\Psi|^2 +
a(T)|\Psi|^2 + b(T)|\Psi|^4 + c(T)|\Psi|^6 +\cdots
\label{e27}
\end{equation}
where the parameters $a(T)$, $b(T)$, $c(T)$, $\ldots$ depend on the
temperature $T$ as well as the detail form of the interaction
Hamiltonian between the helium atoms. Let us first cut off the expansion at
the second order term; we will consider the affects of higher order
terms later. We then have a Landau--Ginzberg type of free energy, and
eq.~$(\ref{e26})$ becomes
\begin{equation}
0=-\frac{1}{2m}\nabla^2\Psi_{\min} +a(T)\Psi_{\min} +
2b(T)|\Psi_{\min}|^2\Psi_{\min},
\label{e28}
\end{equation}
Requiring that $\Psi_{\min}$ be harmonic
\begin{equation}
0=\Psi_{\min}\left(a(T) + 2b(T)|\Psi_{\min}|^2\right),
\label{e29}
\end{equation}
The general solution of eq.~$(\ref{e28})$ is
\begin{equation}
\Psi_{\min}=0 \qquad{\rm or}\qquad \Psi_{\min} = \pm
\left(-\frac{a(T)}{2b(T)}\Bigg|_{\min}\right)^{1/2}.
\label{e30}
\end{equation}

While the field $\Psi$ does not have to be a global function, the
free energy functional must be. Consequently, $|\xi|=1$. From the
Maximum Modulus Theorem, the only such holomorphic function is the
constant function. We will therefore define $\xi$ as the abelian
group generated by $e^{i\Psi_{\min}}$. The exact form of $\xi$, as
well as the phase of $\Psi_{\min}$, was chosen so that $|\xi|=1$.

Vortices exist because of non-trivial differences in the phase of the
macroscopic field $\Psi$ in various regions of the Bose liquid. To
each configuration of vortices there is an open covering
of $\mathbf M$ and a set $\{\xi_{\alpha\beta}\}$ subordinate to the
intersection $U_\alpha\cap U_\beta$. A non-trivial choice, for example
\begin{equation}
\xi_{\alpha\beta} =\left(e^{i\Psi}\right)^3
\label{e31}
\end{equation}
would result in the presence of a vortex with strength $3$ within this
intersection. Going from $U_\alpha$ to $U_\beta$ changes the phase of
$\Psi$ by $\xi_{\alpha\beta}$. The exact value of $\Psi_{\min}$ does
not matter as long as it is non-zero.

In the phenomenological theory given in~\cite{SM} the superfluid
state is marked by the presence of vortices while the normal fluid
state does not have vortices. Consequently, $\xi =1$ when $T>T_c$ and
when $T<T_c$, $\xi\not= 0$. From eq.~$(\ref{e29})$, necessarily
$b(T)>0$ for $T>T_c$, corresponding to $\Psi=0$, while for $T<T_c$,
$a(T)/b(T)<0$. This is the standard result expected from a
Landau--Ginzberg theory, but now it is necessitated by the presence and
role that vortices play in the superfluid transition.

We could have chosen to include a higher order term in the
expansion of~$F$, say to third order. Then it is possible that
other linearly independent solutions could exist for the free
energy minima, and each additional solution would enlarge the
gauge group, although it will still remain abelian. For example,
the gauge group for the quadratic free energy is a global $U(1)$,
for the third order polynomial one could have is $U(1) {\mathrm X}
U(1)$ and so on. However, each time the gauge group is enlarged,
the genus of $\mathbf M$ necessarily increases. Going back to
eq.~$(\ref{e4})$, we see that the hamiltonian admits affine
transformations of the coordinates $\{z_k\}$. The transition
functions of the manifold $\mathbf M$ must thus be affine.
A~compact Riemann surface $\mathbf M$ admits affine
transformations if and only if the chern class of the canonical
bundle vanishes. As $c(\kappa)=2(g-1)$, we find that $\mathbf M$
must be a $1$-torus, From this argument then the only physically
relevant free energy for superfluidity in two dimensions can have
at most one non-trivial root. Any additional terms in $F$ of an
order higher that $|\Psi|^4$ will not introduce any additional
physically relevant solutions at any temperature.

Finally, for $g=0$, dim $H^1({\mathbf M}, {\mathbf C}) =0$ and there are no
lines bundles whose chern class vanishes. Since the total
charge of the system must vanish, we conclude that vortex precipitation
cannot occur on the surface of a sphere. Vortex precipitation can,
however, occur on the surface of a rectangular plane which can be
compactified by identifying opposite sides. In doing so, we have
mapped the plane into a torus, which is exactly the structure
required by the Riemann--Roch theorem.

\section{Summary}
In summary, we have shown that the presence of vortices in the
superfluid state necessitates the existence of an order parameter for the
two dimensional vortex system. Since the net charge of the system is
constrained to be zero, these vortices cannot precipitate on the
surface of a sphere or any other surface homeomorphic to it. This
result has been seen experimentally~\cite{KW} and analysed in detail
theoretically~\cite{Spe}. Furthermore, the symmetries of the
hamiltonian require that the free energy be a functional of at most
one macroscopic field. In addition, for the phase transition to take
place, there can be at most one nontrivial solution of the equations
minimizing~$F$ at any temperature.

\label{Morrison-lastpage}

\end{document}